\documentclass[sigconf, screen]{acmart}
\usepackage{csquotes}
\usepackage{hyperref}
\hypersetup{
    colorlinks=true,
    linkcolor=teal,
    filecolor=teal,      
    urlcolor=teal,
    pdftitle={Overleaf Example},
    pdfpagemode=FullScreen,
}
\MakeOuterQuote{"}
\AtBeginDocument{%
  \providecommand\BibTeX{{%
    \normalfont B\kern-0.5em{\scshape i\kern-0.25em b}\kern-0.8em\TeX}}}

\copyrightyear{2024}
\acmYear{2024}
\setcopyright{rightsretained}
\acmConference[C\&C '24]{Creativity and Cognition}{June 23--26, 2024}{Chicago, IL, USA}
\acmBooktitle{Creativity and Cognition (C\&C '24), June 23--26, 2024, Chicago, IL, USA}
\acmDOI{10.1145/3635636.3656192}
\acmISBN{979-8-4007-0485-7/24/06}

\usepackage{xcolor}
\colorlet{RED}{red}
\newcommand{\rev}[1] {{#1}}
\newcommand{\camera}[1] {{#1}}

\begin{document}

%%
%% The "title" command has an optional parameter,
%% allowing the author to define a "short title" to be used in page headers.
\title{VideoMap: Supporting Video Editing Exploration, Brainstorming, and Prototyping in the Latent Space}

%%
%% The "author" command and its associated commands are used to define
%% the authors and their affiliations.
%% Of note is the shared affiliation of the first two authors, and the
%% "authornote" and "authornotemark" commands
%% used to denote shared contribution to the research.
% \author{Anonymous Author(s)}
\author{David Chuan-En Lin}
\affiliation{%
  \institution{Carnegie Mellon University}
  \streetaddress{5000 Forbes Ave.}
  \city{Pittsburgh, PA}
  \country{USA}
  }
\email{chuanenl@cs.cmu.edu}

\author{Fabian Caba Heilbron}
\affiliation{%
  \institution{Adobe Research}
  \streetaddress{}
  \city{San Jose, CA}
  \country{USA}
  }
\email{caba@adobe.com}

\author{Joon-Young Lee}
\affiliation{%
  \institution{Adobe Research}
  \streetaddress{}
  \city{San Jose, CA}
  \country{USA}
  }
\email{jolee@adobe.com}

\author{Oliver Wang}
\affiliation{%
  \institution{Adobe Research}
  \streetaddress{}
  \city{Seattle, WA}
  \country{USA}
  }
\email{owang@adobe.com}

\author{Nikolas Martelaro}
\affiliation{%
  \institution{Carnegie Mellon University}
  \streetaddress{5000 Forbes Ave.}
  \city{Pittsburgh, PA}
  \country{USA}
  }
\email{nikmart@cmu.edu}

%%
%% By default, the full list of authors will be used in the page
%% headers. Often, this list is too long, and will overlap
%% other information printed in the page headers. This command allows
%% the author to define a more concise list
%% of authors' names for this purpose.
\renewcommand{\shortauthors}{Lin et al.}

%%
%% The abstract is a short summary of the work to be presented in the
%% article.
\begin{abstract}
  Video editing is a creative and complex endeavor and we believe that there is potential for reimagining a new video editing interface to better support the creative and exploratory nature of video editing. We take inspiration from latent space exploration tools that help users find patterns and connections within complex datasets. We present VideoMap, a proof-of-concept video editing interface that operates on video frames projected onto a latent space. We support intuitive navigation through map-inspired navigational elements and facilitate transitioning between different latent spaces through swappable lenses. We built three VideoMap components to support editors in three common video tasks. In a user study with both professionals and non-professionals, editors found that VideoMap helps reduce grunt work, offers a user-friendly experience, provides an inspirational way of editing, and effectively supports the exploratory nature of video editing. We further demonstrate the versatility of VideoMap by implementing three extended applications. \camera{For interactive examples, we invite you to visit our project page: \url{https://humanvideointeraction.github.io/videomap}}.
\end{abstract}

%%
%% The code below is generated by the tool at http://dl.acm.org/ccs.cfm.
%% Please copy and paste the code instead of the example below.
%%

%%
%% Keywords. The author(s) should pick words that accurately describe
%% the work being presented. Separate the keywords with commas.

%% A "teaser" image appears between the author and affiliation
%% information and the body of the document, and typically spans the
%% page.
% \textwidth

\begin{teaserfigure}
  \centering
  \includegraphics[width=15cm]{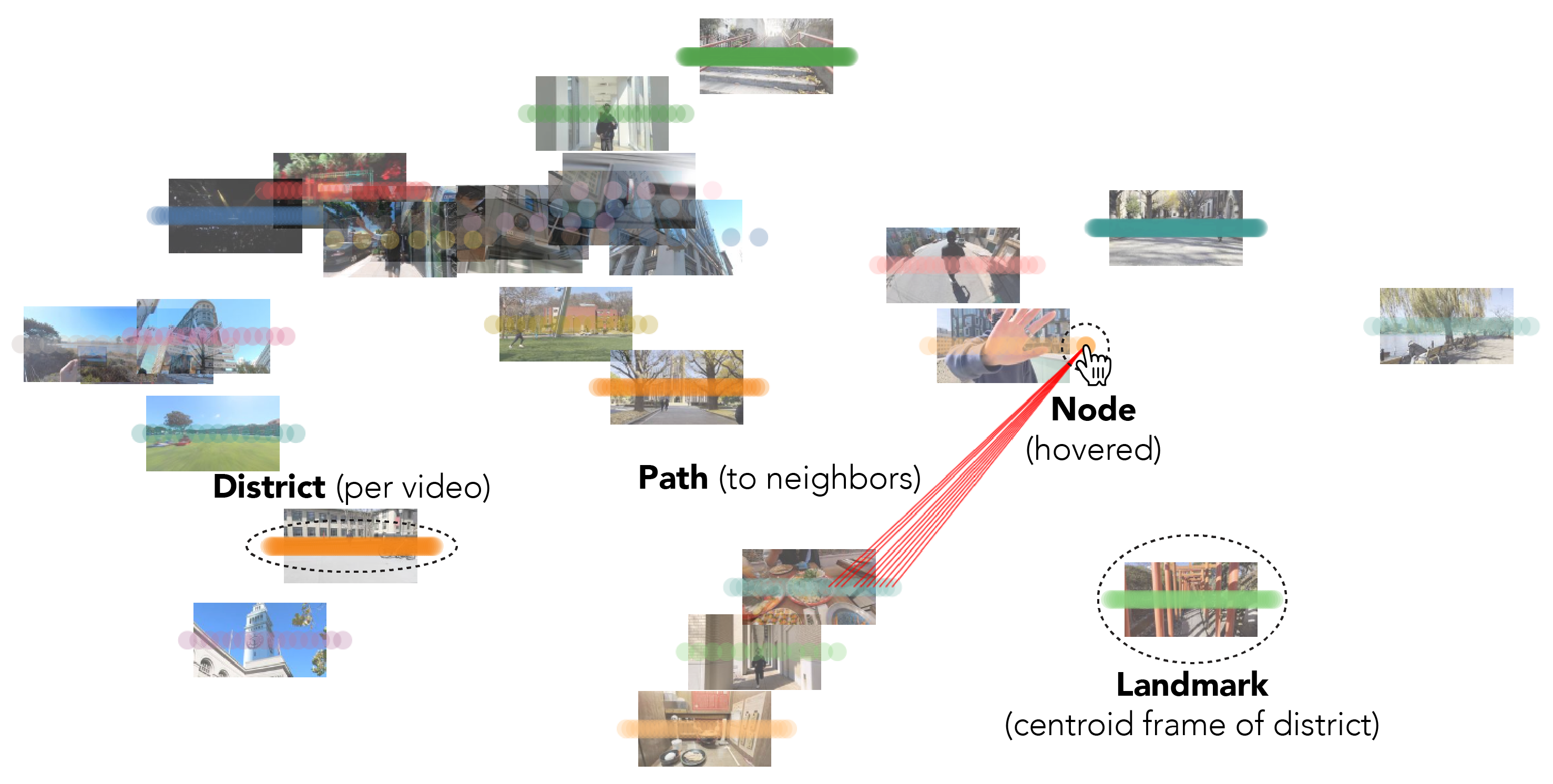}
  \caption{VideoMap is a proof-of-concept video editing interface that operates on video frames projected onto a latent space, enabling users to visually uncover patterns and relationships. We introduce map-inspired navigational elements (node, path, district, landmark) to support users in navigating the map.}
  \Description{VideoMap is a proof-of-concept video editing interface that operates on video frames projected onto a latent space, enabling users to visually uncover patterns and relationships. We introduce map-inspired navigational elements (node, path, district, landmark) to support users in navigating the map.}
  \label{fig:teaser}
\end{teaserfigure}

%%
%% This command processes the author and affiliation and title
%% information and builds the first part of the formatted document.
\maketitle

\section{Introduction}
Video editing is a creative and complex endeavor. Consider a typical workflow: As the editor sifts through large amounts of footage, they must first develop a comprehensive understanding of the material at hand and conceptualize a narrative. The editor then needs to select the individual clips to use and identify connections between them to weave them together with suitable transitions. As the editor refines the edit, they may constantly experiment with different editing ideas.

% \vspace{0.3cm}
% \begin{quotation}
% \textit{``Editing is one of the most challenging and rewarding aspects of filmmaking. It's where you really get to play with time and space and shape the story in a completely unique way.''}
% \vspace{0.1cm}
% \par\raggedleft--- \textup{Steven Spielberg, Film Director}
% \end{quotation}
% \vspace{0.3cm}

We observe that popular video editing interfaces used by video editors often feature a sequential editing timeline, with designs rooted in the metaphors of hand cutting and splicing film \cite{video-editing-history}, supported by a grid-like asset management panel, where clips are listed by filename order or time of creation \cite{premiere-project-panel}.
% remained largely unchanged over the past two decades \cite{casares2002simplifying}. These interfaces 
While these editing interfaces functionally support the task of assembling video frames in a specific order, they \camera{do not have explicit mechanisms to support the \textit{creative exploratory aspects of editing}, such as developing a holistic understanding of all video footage, identifying connections between them, and rapidly experimenting with multiple editing ideas. We believe that there is an opportunity to build a new video editing interface that better supports this creative process. We aim to augment the way editors see and interact with their video clips during the \textit{early stages} of video editing, including exploration, brainstorming, and prototyping.}
% TODO: Think another word for prototyping for videos

In this research, we take inspiration from latent space exploration tools that help users find patterns and connections within complex datasets \cite{liu2019latent}. Latent space exploration tools leverage data processing and recent machine learning techniques to transform data into vector representations and enable users to visually explore the data on a spatial interface.
% Recent work has leverage video-specific autoencoders to map video frames onto latent spaces and suggests that such mapping could be useful for video editing \cite{wang2021video}.
Building upon this insight, we ask if we can similarly project video frames onto a latent space and allow editors to edit videos within this space.

To investigate this possibility, we developed VideoMap, a proof-of-concept video editing interface based on latent space representations of video frames. In our system, we encode video frames onto latent spaces that are meaningful for video editors and support specific video editing tasks. For example, we created a latent space in which video frames containing similar shapes are near each other to help editors identify opportunities to link together scenes with similar shapes and create seamless transitions. Editors may switch between different types of latent space maps that encode information meaningful to video to focus on different types of tasks, a concept we call ``swappable lenses.'' To facilitate more intuitive navigation of the map, we introduce a set of map-inspired navigational elements: nodes, districts, landmarks, and paths \cite{lynch1984reconsidering}. We then designed three components for VideoMap that utilize the latent space map to support three common video tasks. Specifically, we designed 1) Project Panel to help editors explore video footage, 2) Paths Explorer to help editors find suitable video transitions, and 3) Route Planner to help editors quickly prototype rough cuts and try out different editing ideas.

In a user study, we invited both professional and non-professional video editors to test VideoMap. We found that video editors were able to effectively use VideoMap's components to perform the editing tasks listed above. Editors expressed that VideoMap provides a user-friendly editing experience, reduces tedious grunt work, enhances the overview capability of video footage, helps identify suitable video transitions, and enables a more exploratory approach to video editing. We further explored the design space of VideoMap by showcasing how VideoMap can be customized and extended to support additional applications, including summarizing videos, finding highlight moments within videos, and text-based video editing.
% \vfill\break

This research thus makes the following contributions:
\begin{itemize}
\item {We introduce \textbf{VideoMap, a proof-of-concept video editing interface that operates on video frames projected onto a latent space}. We support intuitive navigation of the latent space through map-inspired navigational elements and facilitate transitioning between different latent spaces using swappable lenses.}
\item{We built \textbf{three VideoMap components} to support editors in three common video tasks: exploring video footage, finding suitable video transitions, and quickly prototyping rough cuts.}
\item{We demonstrate the effectiveness of VideoMap in supporting video editors complete creative editing tasks through a \textbf{user study} ($N$=14). Editors felt that VideoMap provides a user-friendly editing experience, reduces tedious grunt work, enhances the overview capability of video footage, promotes editing continuity, and enables a more exploratory approach to video editing.}
\item{We further demonstrate the versatility of VideoMap by implementing \textbf{three extended applications}, showcasing how future developers may customize and extend VideoMap for additional use cases.}
\end{itemize}

\section{Related Work}

Our work is situated among literature on video editing tools, video browsing interfaces, and latent space exploration tools.

\subsection{Video Editing Tools}

VideoMap contributes to extensive literature on enhanced video editing tools. In this section, we discuss some past techniques, such as leveraging video metadata, pen input, tangibles, text-based editing, and automation.

Early research on video editing tools explored how data embedded within and associated with videos could be used to create new editing interfaces. One of the seminal works, SILVER \cite{casares2002simplifying}, uses video metadata to support editors with additional panels to interface with and edit video footage, such as storyboards, editable transcripts, and timeline views. SILVER shows how the use of video metadata allows for smart editing capabilities beyond simple cutting, trimming, and arranging of video frames on the timeline. 
\cite{girgensohn2000semi} analyzes video content for fast motion or zooming and computes a ``suitability'' score for each video frame to help editors decide which clips to include in the edit. This suitability score takes information about the video embedded in the pixels and makes it available as metadata for editing purposes, showing how metadata derived computationally from the frames can be leveraged to support creative editing work.
VideoMap builds upon these works and leverages new kinds of metadata derivable with modern machine learning techniques.

Early researchers have also looked into alternative ways to arrange and trim videos.
\cite{cattelan2008watch} proposes a way to edit video where the user gives voice comments over the video and the system creates a newly edited video based on the comments. 
Other research has explored augmenting video editing using pen-based technology, allowing editors to edit videos with direct manipulation \cite{cabral2012videoink, tong2011video, cabral2017video}. Another thread of research explores using tangible interfaces to allow co-located users to edit videos collaboratively \cite{sokoler2002videotable, zigelbaum2007tangible, merz2018clipworks}.

More recent research have mostly explored developing video editing tools that cater to a specific domain. A large thread of work focuses on building editing tools for text-focused videos, such as voiceover videos and talking head videos. \cite{leake2017computational} helps users edit dialogue-driven scenes by matching video clips to relevant dialogue. \cite{fried2019text} lets users edit talking-head videos by editing a text transcript. \cite{huber2019b} helps recommend b-roll video clips via interactive transcripts. Similarly, \cite{wang2019write} lets users edit text and automatically recommends video clips they filmed to use in the edit. \cite{leake2020generating} helps users create videos by recommending matching images over text and transcripts through the notion of word concreteness (i.e., the extent to which a word describes something that can be visually experienced). \cite{xia2020crosscast} allows users to add images to podcasts by using natural language processing to identify important geographic locations and descriptive keywords. Today, several commercial, text-based video editing tools allow users to edit videos via a paired transcript, including Descript \cite{descript} and Type Studio \cite{typestudio}. In Section \ref{section:text-based-video-editing}, we demonstrate how VideoMap may also be extended to support text-based video editing.

There have also been efforts to automate the video editing process using algorithmic techniques, including identifying highlight segments \cite{hua2003ave, lin2024videogenic, wang2009evolutionary}, mimicking professionally-created videos \cite{pardo2021learning, huang2021learning, huang2019learning,  chen2023match}, \camera{instructing editing objectives in natural language to an LLM-powered agent \cite{wang2024lave}, and automatically synthesizing sound based on video content \cite{lin2023soundify}.} In addition, researchers have also explored automatically converting various forms of content into videos, such as web page to video \cite{chi2020automatic}, markdown to video \cite{chi2021automatic}, and physical demonstrations to video \cite{chi2013democut}.
Although the main objective of VideoMap is not on automatic video editing, we designed a feature that supports editors in automatically generating an initial draft of a video (see Section \ref{section:route-planner}), suggesting its potential to support new kinds of video editing automation.

\subsection{Video Browsing Interfaces}

A related research area to video editing tools focuses on interfaces to support video browsing. Browsing and selecting videos from a large set of raw footage is an important first step in the video editing process. A large body of research investigates the use of direct manipulation techniques \cite{hutchins1985direct} for video browsing. Many works have explored video playback by allowing the user to directly click and drag objects within the video, such as clicking on a car and dragging it along the road to play the video over time \cite{kimber2007trailblazing, karrer2008dragon, nguyen2013direct, goldman2008video}. \cite{pongnumkul2010content} creates a timeline with dynamic playback speeds to emphasize important content. Another thread of research supports users in browsing through videos by revealing additional information, such as additional thumbnails \cite{matejka2013swifter, jackson2013panopticon}, visualizing various types of metadata such as brightness intensity and hue histograms \cite{tonomura1993videomap}, enhancing the playback of software tutorial videos by leveraging crowdsourced data from previous user interactions with the tutorials, \cite{yang2022softvideo}, and improving the playback of surgical videos by surfacing related surgical videos that perform similar surgical steps \cite{kim2023surch}. In VideoMap, we leverage the content embedded within videos to enable enhanced browsing of video clips and visualization of connections between video clips. This allows editors to explore videos based on their content, rather than merely viewing them as a list of files.

In addition to visual browsing methods, researchers have also explored adopting text-based browsing and navigation for content-heavy videos, such as for lecture videos \cite{pavel2014video, yang2014content}, movies \cite{pavel2015sceneskim}, presentation videos \cite{peng2021slidecho}, and how-to videos \cite{chang2021rubyslippers}. Such text-based browsing methods allow editors to quickly find the right moment in the footage, helping them focus more on composing the video than on searching for content. In VideoMap, we also leverage text-based search of features derived from a semantic latent space to allow editors to find content within large amounts of footage (see Section \ref{section:prompts}).

\subsection{Latent Space Exploration Tools}

With advancements in machine learning, researchers in visualization and interpretable machine learning communities have recently explored methods for visualizing the embeddings of neural networks. These embeddings are vector representations of data that have been learned by neural networks. A common initial step of such works involves reducing the dimensions of these vector representations, which are often high-dimensional, to just two dimensions. This allows for the data to be visually plotted using $x$ and $y$ coordinates, resulting in a 2D latent space. Various techniques can be used to reduce the dimension of embeddings while retaining as much information as possible. Some popular methods include t-SNE \cite{van2008visualizing}, HSNE \cite{pezzotti2016hierarchical}, PCA \cite{abdi2010principal}, and UMAP \cite{mcinnes2018umap}. In VideoMap, we adopt dimensionality-reducing preprocessing to develop a new interface for viewing and working with video footage.

Given a 2D latent space, researchers have investigated the development of interactive interfaces to support user exploration \cite{smilkov2016embedding}. \cite{liu2019latent} develops an interface to help users interactively discover meaningful relationships among data points in the latent space. \cite{eckelt2022visual} allows users to discover structural relationships in data, such as the association of items within groups and the hierarchies of items between groups items.
Several works have investigated interfaces for visualizing semantic relationships of text data \cite{liu2017visual, yao2018dynamic}. 
\cite{chang2018recipescape} analyzes cooking processes at scale by clustering recipes with respect to their structural similarities and facilitates pairwise comparisons between recipes.
\cite{dang2021gesturemap} allows users to visualize a latent space of gestures. \cite{grossmann2022concept} support users in visualizing a latent space of human interpretable concepts, such as cars.
Many researchers have also worked on tools for comparing multiple latent spaces, such as by highlighting visual changes \cite{sivaraman2022emblaze} and through side-by-side comparisons \cite{li2018embeddingvis, boggust2022embedding}.
Most relevant to our work, \cite{wang2021video} train video-specific autoencoders (i.e., autoencoder models overfitted to the task of regenerating the videos) to visualize video frames on a latent space, which could be used to perform video manipulations such as inpainting and creating video textures (i.e., looping videos).
In VideoMap, we draw inspiration from latent space visualization interfaces to assist video editors in identifying patterns and connections within video footage.
We create latent spaces that are meaningful for video editors, representing various aspects of video editing, such as concepts of semantics, color, and shape.

% Specifically, we build upon Wang, Ramanan, \& Bansal \cite{wang2021video} who show simple examples of editing videos by finding loops among frames mapped onto the latent space or by allowing users to create a path between frames to create short video clips.
% In our work, we extend these simple editing concepts to support more professional, creative aspects of editing across video clips and through latent spaces representing data that is more meaningful to editors, such as semantics, color, and shapes.

\begin{figure*}[!htb]
  \centering
  \includegraphics[width=15cm]{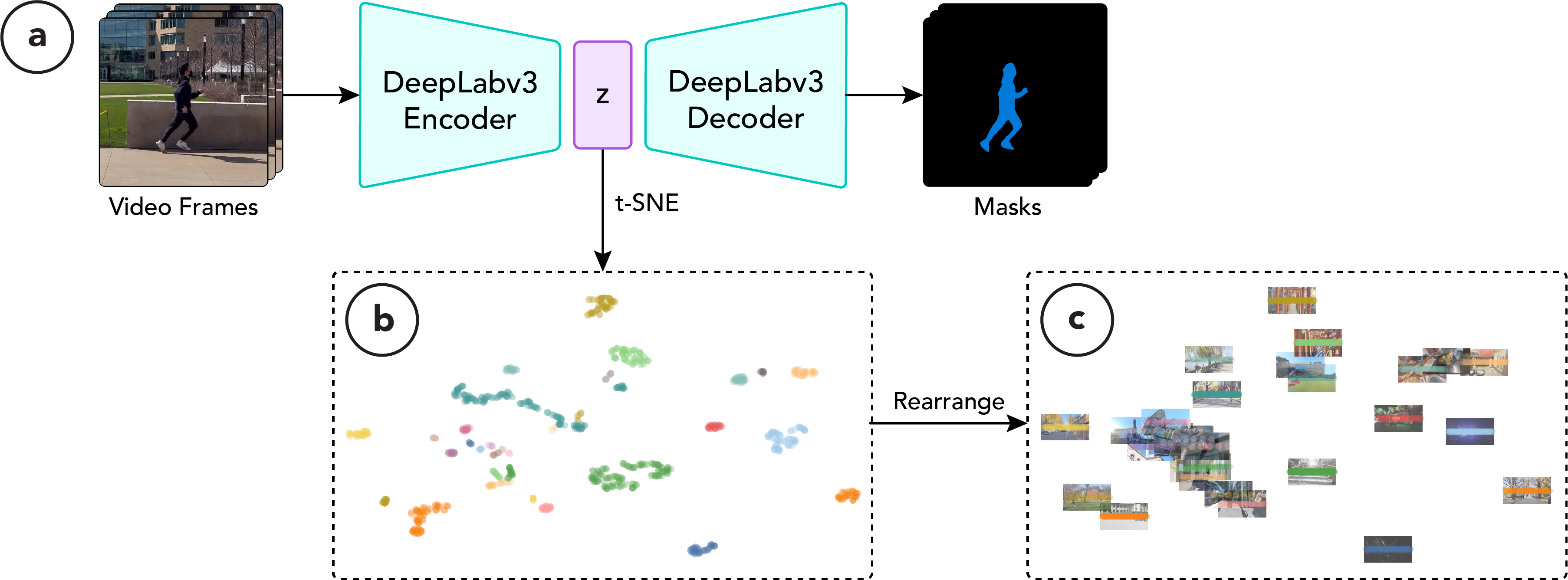}
  \caption{The pipeline for creating the \texttt{shape} lens. We pass video frames through DeepLabv3, an image masking model, to extract shape-related vectors (a). We then apply t-SNE to reduce the vectors to two dimensions (b) and visually rearrange the vectors by video to make the videos more skimmable (c).}
  \Description{The pipeline for creating the \texttt{shape} lens. We pass video frames through DeepLabv3, an image masking model, to extract shape-related vectors (a). We then apply t-SNE to reduce the vectors to two dimensions (b) and visually rearrange the vectors by video to make the videos more skimmable (c).}
  \label{fig:shape-pipeline}
\end{figure*}

\section{Principles of VideoMap}
We distill three principles to ground the development of VideoMap based on theories and practices in video editing. These guiding principles include providing an overview of the video footage, maintaining editing continuity with seamless transitions between clips, and facilitating exploration and experimentation throughout the editing process.

\subsection{Principle 1: Overview}
As a time-based medium, video editing often requires editors to spend a considerable amount of time sequentially playing through and reviewing footage on an editing timeline to develop an understanding of the material they are working with. In VideoMap, we aim to overcome the sequential nature of browsing through videos by projecting video clips onto a 2D latent space, which offers a more comprehensive overview of the footage at a glance. Findings from Craft and Cairns \cite{craft2005beyond} suggest the patterns and themes in data may only be seen from a vantage point that comprises the whole view. 
Map-like interfaces, such as plots on a 2D latent space, can provide such an overview of data and have been shown to allow people to better understand the thematic information across a dataset \cite{hografer2020state}.
We posit that with an enhanced overview, editors will be able to more easily identify patterns and structure and conceptualize a narrative thread from a set of video clips. 
Furthermore, we designed ``swappable lenses'' for editors to easily switch between different types of overviews (i.e., different latent spaces) to uncover different types of patterns within the footage, such as patterns in semantics, color, or shape (see Section \ref{section:swappable-lenses}). We designed the Project Panel component to test how VideoMap can support editors through an enhanced overview capability.

\subsection{Principle 2: Continuity}
Continuity is one of the key principles of video editing. In the 2012 Asia-Pacific Symposium on Creative Post-Production, Richard Crittenden, author of the book Film and Video Editing, comments: ``\textit{It remains true that good editing tends to be the art that conceals art}''. The goal of continuity is to connect clips seamlessly so that the edits are ``invisible'' to the viewer \cite{reisz1971technique}. One editing technique that editors use to achieve continuity is called match cutting \cite{match-cuts}. For example, the editor could cut from a video frame of a bagel to a video frame of a donut — using two different objects with similar shapes to transition between present and past scenes, for example. In VideoMap, we aim to assist editors in achieving continuity by recommending seamless video transitions. Since VideoMap operates in the latent space, we suggest video transitions for a video clip or frame by finding near neighbors in the latent space. Taking the shape-based match cut example: By creating a latent space where video frames containing similar shapes are organized near each other, the bagel video frame and the donut video frame will be near neighbors in the latent space. We designed the Paths Explorer component to test how VideoMap can support editors in achieving continuity in editing.

\subsection{Principle 3: Exploration}
Even with a well-defined narrative and a principle of continuity to follow, video editing remains a process that involves exploration and experimentation. Film director Ridley Scott likened editing to the process of painting: ``\textit{Editing is like painting, only with film. You are constantly making choices and trying to find the best way to tell the story.}'' In VideoMap, we aim to provide editors with a wide range of exploration mechanisms. First, we created four map-inspired navigational elements to help editors explore the latent space: nodes, paths, districts, and landmarks \cite{lynch1984reconsidering}. Second, we implemented a semantic searching mechanism in the Project Panel component, allowing editors to quickly locate and discover clips using text prompts. Third, we recommend multiple alternative video transition options for editors in the Paths Explorer component. Finally, we designed the Route Planner component, which can automatically assemble video cuts from user-selected clips, enabling editors to quickly test and preview the results of different editing ideas.

\section{Implementation}
\label{section:implementation}
Our three principles are manifested in VideoMap and guide its implementation. The following outlines the implementation of our system. We first detail how we built the base layer latent space map that powers the VideoMap interface. We then go through how we designed VideoMap's main components, including the Project Panel for exploring video footage, Paths Explorer for finding suitable video transitions, and Route Planner for prototyping video route cuts.

\subsection{Building the Map}
We implemented VideoMap's base layer by first creating the 2D latent space. We then enable flexible reconfiguration of different latent spaces using swappable lenses, and support intuitive navigation of the map through map-inspired metaphors.

\subsubsection{Creating the Latent Space}
\label{section:creating-the-latent-space}

VideoMap is based on a latent space representation of video frames. Specifically, we encode video frames into vector representations such that they are meaningful for particular video editing tasks. For example, we can create a latent space map to help editors find shape match cuts, a type of video transition which links together two scenes that contain similar shapes to create a smooth transition (\href{https://www.youtube.com/watch?v=W2hjlA1rEfM}{see an example from Kubrick's 2001: A Space Odyssey}). To create this map, we pass video frames through DeepLabv3 \cite{chen2017rethinking}, an image masking model, with a ResNet \cite{he2016deep} backbone and extract a 512-dimensional vector representation right after the last visual layer (Figure \ref{fig:shape-pipeline}a) Our insight is that due to the mask generation objective, which extracts the dominant shape(s) of an image, vector representations extracted from DeepLabv3 implicitly encode valuable shape-related information. In other words, video frames that contain similar shapes should have more similar vector representations.

\begin{figure*}[!htb]
  \centering
  \includegraphics[width=15cm]{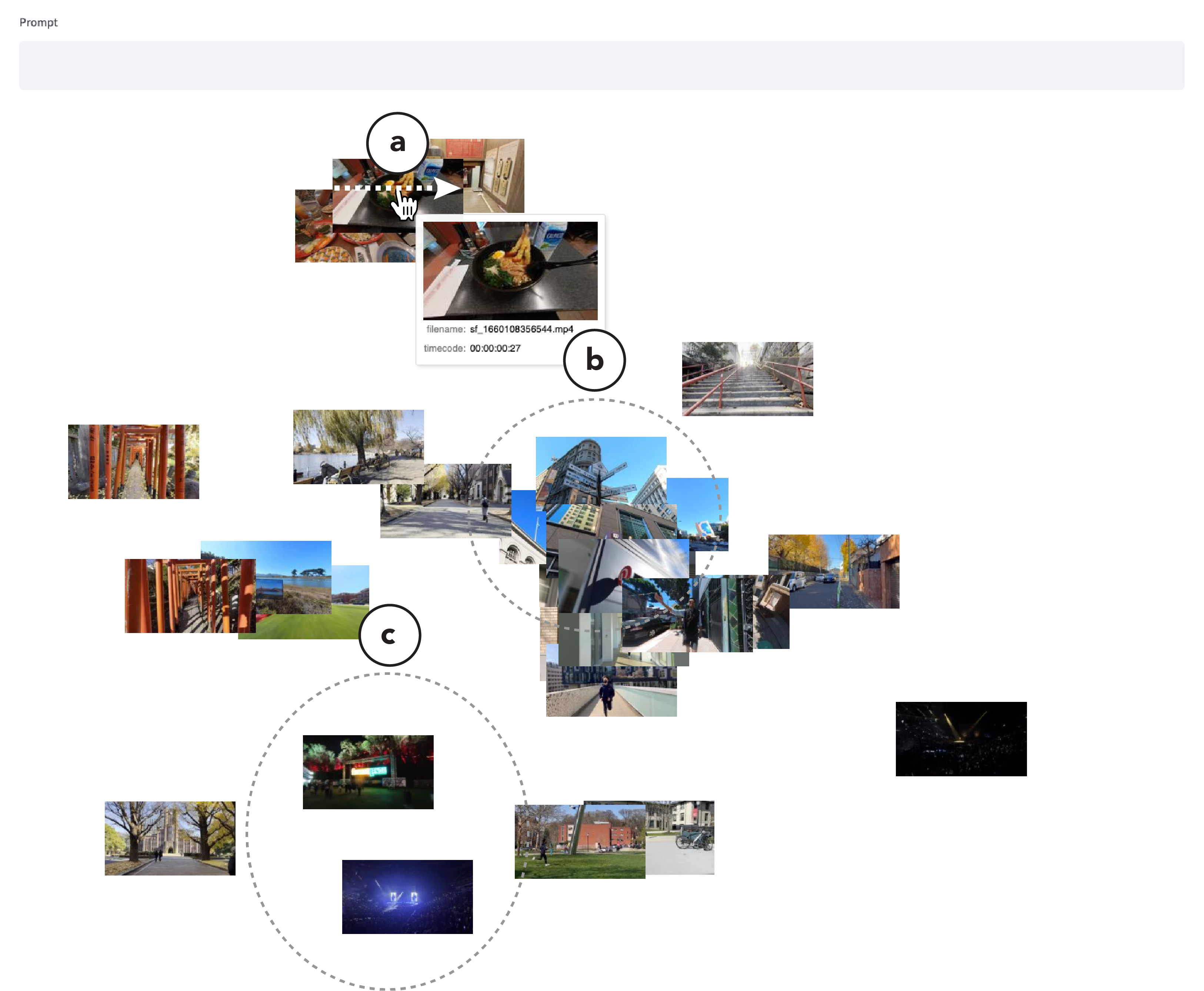}
  \caption{VideoMap's Project Panel component. The figure shows a collection of videos organized under the \texttt{semantic} lens. The editor can play through the videos by scrubbing the landmarks from left to right (a). Example semantic clusters of videos are shown in (b) for several videos containing streets and buildings and (c) for concert videos.}
  \Description{VideoMap's Project Panel component. The figure shows a collection of videos organized under the \texttt{semantic} lens. The editor can play through the videos by scrubbing the landmarks from left to right (a). Example semantic clusters of videos are shown in (b) for several videos containing streets and buildings and (c) for concert videos.}
  \label{fig:project-panel}
\end{figure*}

Next, we project the vector representations of all video frames into a multi-dimensional latent space. To allow editors to browse through the latent space, we apply t-Distributed Stochastic Neighbor Embedding (t-SNE) \cite{van2008visualizing} to reduce the latent space to two dimensions. In the first iteration of VideoMap, we visualized the vector embeddings directly according to the coordinates generated from the t-SNE (Figure \ref{fig:shape-pipeline}b). However, in our pilot tests, editors felt that this was too messy. They also wished to discern groups of points that belong to the same videos more easily and be able to ``play through'' the videos. Therefore, in the current iteration of VideoMap, we visually rearrange the points on the map by video (Figure \ref{fig:shape-pipeline}c). We first compute the centroid of all points for each video, then we then visualize frames as points horizontally across the centroid at equal distances. We also display the centroid video frames. Editors can thus play through each video by scrubbing through the points horizontally. From our user study, we found that editors considered this design to be easy to use (see Section \ref{section:discussion}).

\subsubsection{Developing Swappable Lenses}
\label{section:swappable-lenses}

\begin{figure*}[!htb]
  \centering
  \includegraphics[width=15cm]{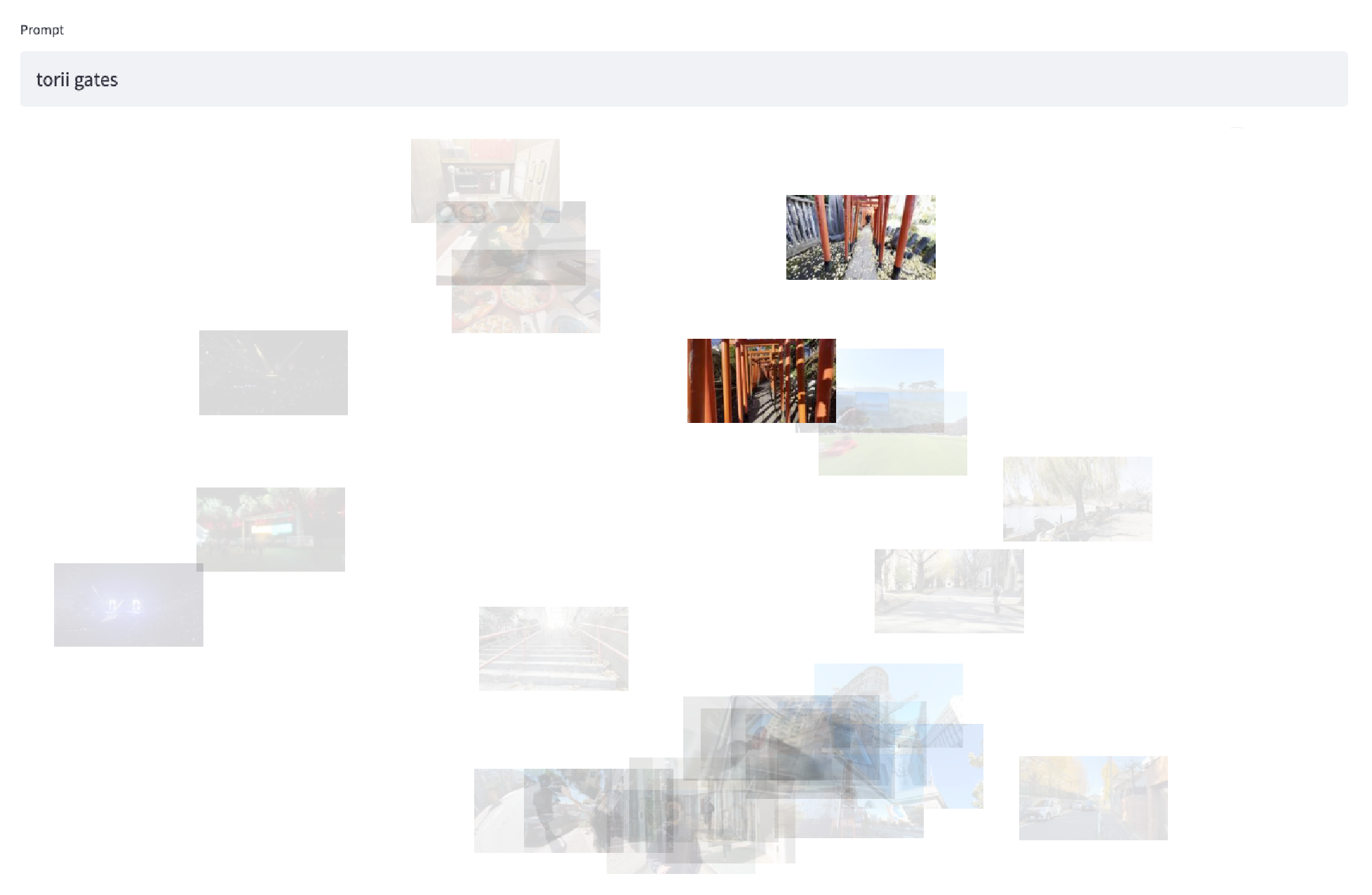}
  \caption{The editor can search for videos using text prompts (e.g., \texttt{torii gates}).}
  \Description{The editor can search for videos using text prompts (e.g., \texttt{torii gates}).}
  \label{fig:prompts}
\end{figure*}

We call the different methods of encoding video frames into meaningful vector representations as different \textit{lenses}. In Section \ref{section:creating-the-latent-space}, we create a \texttt{shape} lens. Under the \texttt{shape} lens, video editors can identify shape-based match cut opportunities by looking at neighbors on the latent space. Similarly, one may develop new lenses by finding other ways of encoding useful vector representations for other use cases. In our proof-of-concept, we implemented a \texttt{color} lens (video frames organized by color) and a \texttt{semantic} lens (video frames organized by meaning). To create the \texttt{color} lens, we first extract 3D histograms from each RGB video frame using 8 bins per channel and normalizing with $range$ = [0, 256], then flatten the histogram to generate a 512-dimensional color-based vector. To create the \texttt{semantic} lens, we encode each video frame through the image encoder of CLIP \cite{radford2021learning}, a neural network that has learned semantic concepts, to generate a 512-dimensional semantics-based vector. Rather than creating a one-size-fits-all map, we allow editors to flexibly swap between different lenses and reconfigure VideoMap according to different criteria. We demonstrate examples of how video editors creatively made use of the different lenses for different editing tasks in Section \ref{section:swappable-lenses-results}.

\subsubsection{Navigating with Nodes, Paths, Districts, and Landmarks}

We designed four map-inspired navigational elements to help editors navigate through the map (Figure \ref{fig:teaser}). The four elements are nodes, paths, districts, and landmarks. These elements are inspired by Lynch's seminal The Image of the City \cite{lynch1984reconsidering}, which studied the elements people use to form mental maps of a city.

\textbf{Nodes.} The node is a point (i.e., a vector representation of a video frame) that the user hovers over with their cursor. We follow Schneiderman's details-on-demand mantra \cite{shneiderman2003eyes} by expanding the node's detailed information, such as its corresponding video frame, video filename, and timecode.

\textbf{Paths.} Paths are lines that connect from a user-selected node to its neighboring points on the latent space. When a user clicks on a node, we display ten paths that start from the node and connect to the ten closest neighboring points, which are the points with the smallest cosine distances to the node. We exclude neighboring points that belong to the same video as the node to ensure that the paths represent video transitions. To obtain an accurate representation of neighbors, we calculate the distances based on the original vector representations rather than the rearranged coordinates on the visual map (see Section \ref{section:creating-the-latent-space}).

\textbf{Districts.} Districts are groups of points with common characteristics. Throughout most of this paper, we present the version of VideoMap where points that belong to the same video are grouped into a district. Note that paths connect across two different districts (i.e., video transitions). We color-code the points by district to visually separate them. We also consider alternative methods for defining districts, such as through groupings of semantically similar video frames, in Section \ref{section:video-summarization}.

\textbf{Landmarks}
Landmarks are reference markers on the map. With landmarks, the user can obtain an overview of the map at a glance (Principle 1). We visualize the corresponding video frames of some points on the map as landmarks. Throughout most of this paper, we present the version of VideoMap where we generate one landmark for each district by visualizing the video frame of the point closest to the district's centroid. We also explore the option of allowing users to define custom landmarks in Section \ref{section:video-highlights}.

\subsection{Project Panel}
\label{section:project-panel}

The Project Panel is the first component that the editor can use on VideoMap (Figure \ref{fig:project-panel}). We designed Project Panel to enable editors to explore video footage more easily (Principle 1). Traditional video editing software typically feature a grid-like interface that displays all of the editor's video files ordered by filename or time of creation. In VideoMap's Project Panel, instead of displaying a standard grid, we rearrange the display of the video files based on the selected lens. For example, Figure \ref{fig:project-panel} displays a collection of videos in the Project Panel under the \texttt{semantic} lens. The editor can swap between different lenses to rearrange the layout and discover new perspectives on the video footage. To reproduce the video thumbnails seen in traditional video editing software, we display the landmark for each video file. The editor may play through the video files by scrubbing the landmarks from left to right. The editor may also zoom in or out of the map by scrolling to view regions of the map with less clutter.

\subsubsection{Prompts}
\label{section:prompts}

The editor can filter videos with text prompts (Principle 3) (Figure \ref{fig:prompts}). For example, by typing \texttt{torii gates}, the editor can highlight these videos and fade out the rest. We encode the text prompt using a CLIP text encoder and compare the text embedding to the image embeddings of video frames, encoded by a CLIP image encoder (i.e., the vector representations under the \texttt{semantic} lens). From our user study, we found that editors felt that this method handled prompt searches effectively (see Section \ref{section:prompts-results}).

\begin{figure*}[!htb]
  \centering
  \includegraphics[width=13cm]{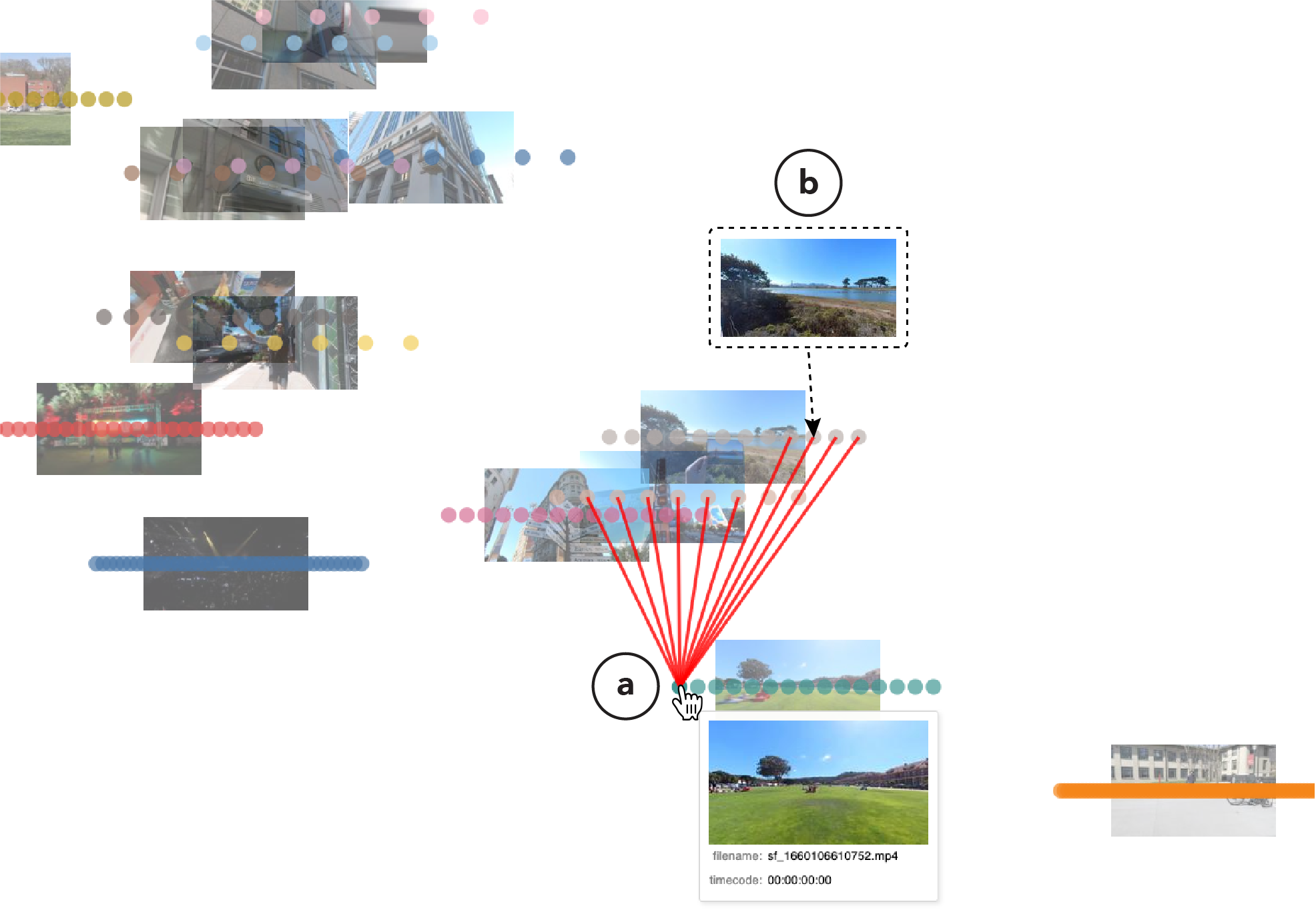}
  \caption{VideoMap's Paths Explorer component. The editor can select a video frame to display ten video transition suggestions based on the selected lens (a). For example, a video frame with a similar color composition is recommended under the \texttt{color} lens (b).}
  \Description{VideoMap's Paths Explorer component. The editor can select a video frame to display ten video transition suggestions based on the selected lens (a). For example, a video frame with a similar color composition is recommended under the \texttt{color} lens (b).}
  \label{fig:paths-explorer}
\end{figure*}

\subsection{Paths Explorer}
\label{section:paths-explorer}

\begin{figure*}[!htb]
  \centering
  \includegraphics[width=16cm]{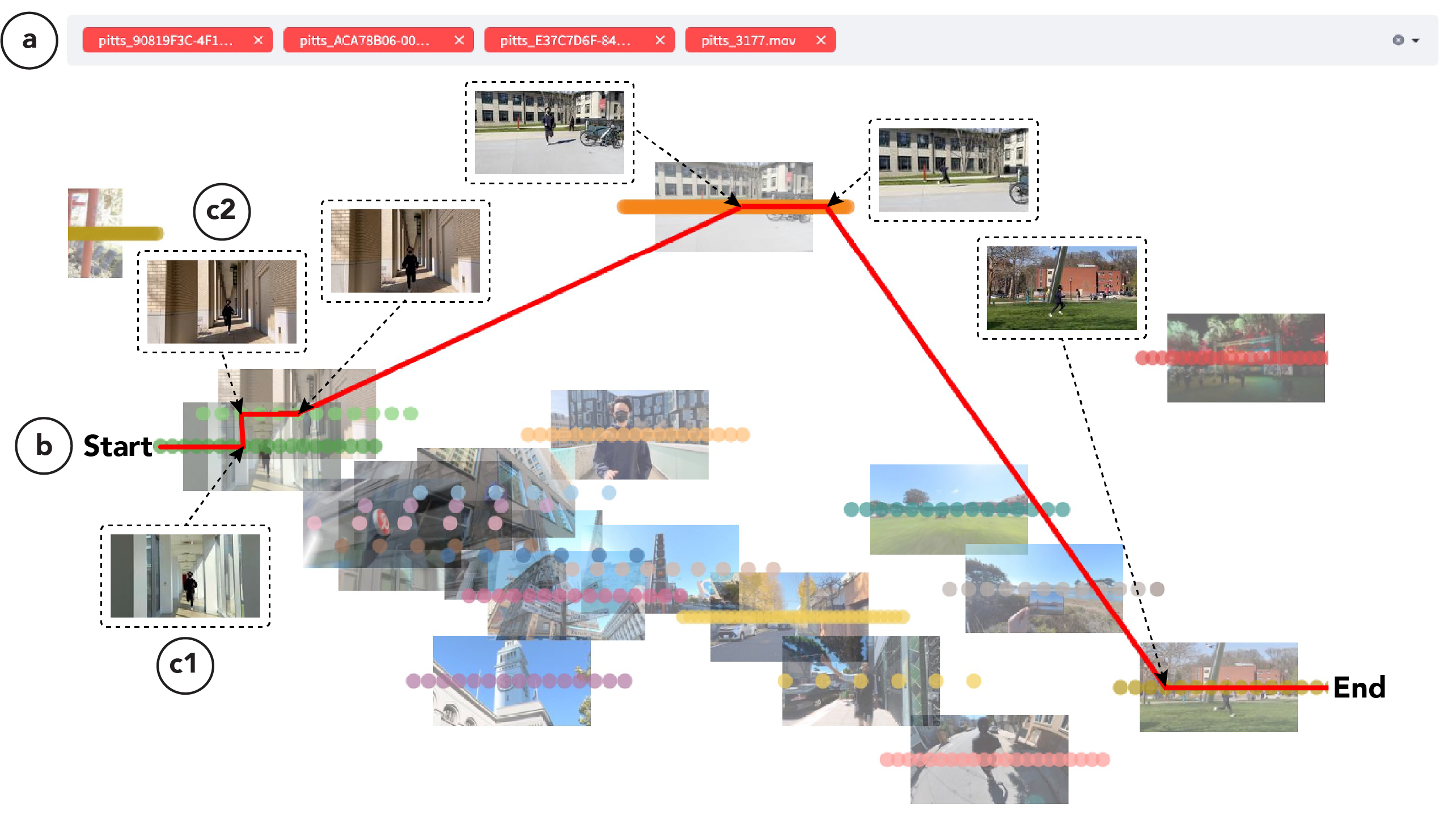}
  \caption{VideoMap's Route Planner component. The editor can select several video clips to automatically generate a rough cut video based on the selected lens (a). For example, the editor can create a video of a person running through various backgrounds under the \texttt{shape} lens (b). Route Planner automatically finds the optimal video transitions (c1 and c2).}
  \Description{VideoMap's Route Planner component. The editor can select several video clips to automatically generate a rough cut video based on the selected lens (a). For example, the editor can create a video of a person running with various backgrounds under the \texttt{shape} lens (b). Route Planner automatically finds the optimal video transitions (c1 and c2).}
  \label{fig:route-planner}
\end{figure*}

Paths Explorer is the second component that the editor can use on VideoMap (Figure \ref{fig:paths-explorer}). We designed Paths Explorer to assist editors in finding suitable video transitions (Principle 2). Editors typically arrange videos in a sequential timeline, which can make it difficult for editors to identify patterns of video frames at different times and across different videos. With Paths Explorer, we overcome the inherent sequential nature of videos by arranging them on a 2D map, thereby removing the limitations of viewing video frames along a 1D timeline (Principle 1). While we only display each video's landmark in Project Panel, we display each video's individual video frames as points in Paths Explorer. We color-code the points by video (districts) and fade the landmarks to reduce visual clutter. When the editor clicks on a point, the interface draws 10 paths connecting to nearest neighbor vector representations from other video(s) (Figure \ref{fig:paths-explorer}b). These paths represent potential video transitions that the system suggests based on the selected lens. For example, Figure \ref{fig:paths-explorer} displays recommended paths that connect between videos with similar color composition (having two prominent blocks of color divided at the horizon) under the \texttt{color} lens. The editor can switch between the lenses to discover different types of video transitions.

\subsection{Route Planner}
\label{section:route-planner}

Route Planner is the third component that the editor can use on VideoMap (Figure \ref{fig:route-planner}. We designed Route Planner to help editors automatically generate rough cuts, which is an initial rudimentary edited version of a video, to quickly explore different video edits (Principle 3). With Route Planner, the editor can select several video clips and the system will automatically find an optimal route across paths (see Section \ref{section:paths-explorer}) that connect these video clips, based on the selected lens. For example, Figure \ref{fig:route-planner} displays an example route that connects across multiple videos containing the same running action under the \texttt{shape} lens. One can think of this component as being analogous to the Google Maps route planner \cite{google-maps}. The paths connecting video clips are like ``streets''. Route Planner finds the fastest route that travels through the streets and passes through each of the selected video clips once. Since the ``distances'' between video clips on the map are determined based on the selected lens, the editor can switch between the lenses to try out different rough cuts based on different criteria, such as switching to \texttt{color} to connect clips with similar color grading or \texttt{semantics} to connect clips depicting similar objects or activities.

To implement Route Planner, we first find the shortest path for all combinations of video pairs (i.e., optimal video transitions). These paths serve as the ``streets'' that connect across different videos. Next, we find the shortest route along the streets that passes through all videos (i.e., the Shortest Hamiltonian Path Problem) with a dynamic programming approach. Finally, we visualize the route on the interface and render the rough cut video by traversing through the route. Since the streets are undirected, some videos may be reversed during the traversal. This can lead to interesting results where videos are connected with one playing forward and one playing in reverse.

\section{User Study}

We conducted a user study to observe how professional and non-professional video editors might use VideoMap to complete creative editing tasks, including developing an overview of the video clips they have, finding unique transitions and cuts between clips, and crafting a rough cut of a video. We aimed to address three research questions with the study:

\begin{itemize}
    \item [RQ1.] How can VideoMap assist editors in exploring video footage?
    \item [RQ2.] How can VideoMap assist editors in brainstorming suitable video transitions?
    \item [RQ3.] How can VideoMap assist editors in rapidly prototyping rough cuts?
\end{itemize}

\subsection{Study Design}
To address our three research questions, we asked video editors to perform the three video editing tasks using VideoMap. Specifically, we asked participants to use the Project Panel to explore and familiarize themselves with a collection of video footage, use Paths Explorer to find video transitions, and use Route Planner to create rough cuts. During the study, we asked participants to share their computer screens and perform a think-aloud \cite{lewis1982using} describing what they were doing and thinking while using VideoMap.

We collected data using three methods. First, we asked participants to complete a questionnaire where participants provided written descriptions on completing the tasks (e.g., a description of how the video clips are organized) as well as feedback on their experience using VideoMap in contrast to how they would perform the same task using a video editing software they are familiar with (see Questionnaire in Supporting   Documents). Additionally, participants attached screenshots of the interface (e.g., a screenshot of a video transition shown in Paths Explorer that they found interesting) and video outputs from completing the tasks (e.g., a rough cut video generated by Route Planner). Second, we recorded the participants' computer screens and transcribed their spoken audio throughout the sessions. Third, the first author silently observed participants working  and noted down interesting insights, such as participants' thoughts while using the system, interaction observations from the screen captures, and the creations participants made while completing the tasks. Overall, the data collected allowed us to gain rich insights into the participants' experiences using VideoMap, as well as visual evidence of how VideoMap was used and could support editors in accomplishing fundamental video editing tasks.

\subsubsection{Video Collection}

We gathered a sample of video clips for participants to use during the editing tasks in our studies. This collection comprises 30 diverse videos, filmed by the authors in various locations including San Francisco, Pittsburgh, and Tokyo. The content covers a variety of categories, including objects like food and architecture, scenery like streets and grassy fields, and actions like running and scootering. The videos were captured using different camera equipment, such as mirrorless cameras, smartphones, and 360 action cameras. \camera{The video clip lengths vary from a couple of seconds to several minutes long.} Figure \ref{fig:project-panel} provides an overview of the sample video collection displayed in VideoMap's Project Panel.

\subsection{Participants}

We recruited 14 participants (8 male, 6 female) aged from 23 to 55 ($\mu$=28.21, $\sigma$=8.78). Seven of the participants are professional video editors recruited from Upwork \cite{upwork}, a platform for hiring creative freelance workers, and seven of the participants are non-professional video editors recruited at our institution via word-of-mouth and snowball sampling. We recruited both professional and non-professional individuals to test VideoMap's ability to support professional editing work as well as make video editing more approachable for non-professionals. We conducted a background survey with the participants before each study to assess their video editing experience. Professional participants have high self-rated familiarity with video editing ($\mu$=5.86, $\sigma$=1.46) (7-point Likert scale) and many years of experience ($\mu$=8.07, $\sigma$=2.55). Non-professional participants have moderate self-rated familiarity with video editing ($\mu$=4.43, $\sigma$=1.27) and several years of experience ($\mu$=3.00, $\sigma$=1.15). Participants use various video editing software such as Adobe Premiere Pro, Final Cut Pro, Movavi Video Suite, iMovie, CapCut, and Google Photos Movie Editor.

\subsection{Procedure}

We conducted the studies over video conference (Zoom). We asked participants to try out the three components of VideoMap (Project Panel, Paths Explorer, and Route Planner) and complete a video editing task using each component.
% Throughout the studies, we asked participants to share their computer screens and perform think-alouds.
The following details our step-by-step procedures.

After obtaining the participant's consent, we collected background information about their prior video editing experience. Next, the participant tried out VideoMap. First, they used the Project Panel to familiarize themselves with the sample video collection and think about how they could create a video montage using it. We also asked participants to try the different lenses (semantic, color, shape) and try searching for clips by typing different prompts in the prompts box. We asked the participant to write down three prompts that yielded interesting results. Second, the participant used Paths Explorer to find video transition opportunities. We asked the participant to record three transitions that they found interesting during their exploration. Third, the participant used the Route Planner to quickly create rough cuts for a video montage using the provided clips. Participants used the Project Panel to determine which video clips to select in the Route Planner for the rough cut. We asked the participant to record one rough cut that they found interesting. After trying out the three components of VideoMap, we asked the participant to complete a post-study written survey about their experiences. We also asked participants to contrast their experiences with how they would typically perform the same tasks using existing video editing software. The study lasted for approximately 1 hour.

\subsection{Results}
\label{section:results}

We analyzed our data, including observation notes, audio transcripts, and form responses, with inductive coding to identify common themes. The following presents our findings. We note direct quotes from professional participants as P and non-professionals as N.

\subsubsection{Project Panel}

Participants were able to use Project Panel to explore the video footage by familiarizing, grouping, and uncovering structure among the clips (RQ1). In addition, participants also noted additional use cases of Project Panel, such as supporting visual ideation and helping with planning new shots to take. As part of the first task, we asked participants to brainstorm narrative ideas that utilize the sample video collection. Some ideas from participants include a video starting from wide-angle shots and then showing details of single activities, a video that first shows campus life and then has a fancy transition to life during the holidays (traveling, concerts), and a video introducing various Japanese cuisines.

\textbf{Familiarize with video footage.} Participants were able to use Project Panel to familiarize themselves quickly with the video collection. P3 expressed that ``\textit{the beginning process of editing a video can be really daunting}'' and Project Panel ``\textit{takes a lot of the time out of needing to familiarize yourself with the content that you have}''. N3 enjoyed using Project Panel as a ``\textit{good preprocessing step}'' and expressed that it ``\textit{can save people time trying to eyeball videos manually themselves}.'' Participants felt that Project Panel ``\textit{corresponds well with [their] mental model of how [they] would look for footage (e.g., grouping common themes)} (N4)'' and facilitates a ``\textit{faster and more intuitive way to explore a collection of video assets} (P4)''.

\textbf{Organizing large projects.} Participants noted that Project Panel could be especially useful for organizing large projects, such as being ``\textit{a unified hub for large projects involving many team members} (P1)''. In particular, many participants enjoyed that Project Panel helps save time and effort required to manually assign tags to video clips or bin video clips into folders (P1, P3, P5, P6, P7, N7): ``\textit{It takes forever to tag clips. It's either that or subfolders. It's such a pain to go through a whole day's worth of shooting}. (P7)''

\textbf{Uncover structure.} Participants stated that Project Panel helps ``\textit{reveal underlying structure within the recorded content} (P4)'', especially those that ``\textit{may not be entirely apparent at first}'' (P4). Participants enjoyed the ``\textit{inspiration and surprise factor} (N5)'' and helps create ``\textit{a fresh perspective on clips they might have passed up} (P7)'': ``\textit{Getting a new take or generating different ideas is what AI is great for and is how we can use tools like these to enhance our editing.} (P7)'' Participants were also able to use Project Panel to easily identify outlier videos (P2, N5): ``\textit{This one color is definitely sticking out} (P2)''.

\textbf{Support visual ideation.} Participants expressed that having the videos visually arranged in a 2D space helped them ``\textit{have a better sense of the clips that they are working with and see the overall vision of what they want to create} (P5)'' (Principle 1). For example, some participants wanted to create a video that takes one scene from each cluster on the map (P1, P5) while others wanted to create a video that focused only on one cluster on the map (P6, N2). N1 noted that ``\textit{seeing similar footage organized in the same place helps [them] figure out where to look if [they are] searching for certain elements} (N1)''.

\textbf{Plan the shots.} Participants reported that Project Panel helped them with planning the (additional) shots that need to be taken (P5, N2). P5 stated that the ``\textit{biggest pro [of Project Panel] was seeing the overall visuals of clips and the amount per grouping. I would use this at the very beginning of my workflow where I decide what other clips I need to take to complete the story. For example, if each scene of a trip is in there or if I need more concert clips.}''

\subsubsection{Prompts}
\label{section:prompts-results}

Participants were able to make use of the prompt box in Project Panel for various purposes, such as quickly searching for clips, searching for similar clips, and searching for cinematic elements. As part of the first task, we asked participants to write down three prompts they found the most interesting during their explorations. The average length of the prompts is around 2 words ($\mu$=2.24, $\sigma$=2.00). The most popular prompt was food, which was included by 5 participants. We grouped the prompts into 6 categories. Overall, participants searched for objects 13 times (e.g., \texttt{bike}), scenery 10 times (e.g., \texttt{outside in a grassy field}), location 6 times (e.g., \texttt{restaurant}), action 7 times (e.g., \texttt{cooking}), time 3 times (e.g., \texttt{daytime}), and cinematography elements 3 times (e.g., \texttt{establishing shot}). Some example prompts from participants include \texttt{videos of people} (P2), \texttt{running} (N2), \texttt{blue sky} (P6), \texttt{autumn colors} (N5), \texttt{motion blur} (N5), and \texttt{establishing shot} (P3).

\textbf{Search for a clip (instantly).} Participants used the prompt box to search for clips quickly: ``\textit{I would be editing at the hotel after a day of filming. Suppose I went to Rome today and I remember I went to the Trevi fountain. It'll be way easier to find it [by searching ‘trevi fountain'].} (P2)'' P5 commented that they could use the prompt box to more easily search for ``\textit{clips with triangular elements while not remembering where those clips were}'' and P3 commented that they could use it to ``\textit{find good b-roll shots}'' for interview-style videos.

\textbf{Search for similar clips.} P7 commented that they could use the prompt box to search for similar clips: ``\textit{There's a concert clip. I wonder if there are any other concert-related clips.}''

\textbf{Search for cinematic elements.} P3 and P4 commented that they could use it to find specific cinematic elements to use for their videos, such as an ``\textit{establishing shot to start off a video} (P3)'' or a ``\textit{wide-angle shot to sprinkle in some drone footage} (P4)''.

\subsubsection{Swappable Lenses}
\label{section:swappable-lenses-results}

Participants found the Swappable Lenses concept useful and expressed that it helped shine a light on different perspectives and allowed VideoMap to be able to cater to people with different editing styles. Participants came up with many creative ways of utilizing Swappable Lenses. P4 used the \texttt{color} lens to find videos of similar seasons, such as autumn. P5 used the \texttt{color} lens to arrange videos chronologically, starting with the day and ending with night scenes. P7 used the \texttt{color} lens to identify videos with warmer and cooler tones. N5 used the \texttt{color} lens to identify black frames to serve as transition points. N6 expressed the potential of using the \texttt{color} lens to maintain a steady color grading scheme. P7 used the \texttt{shape} lens to find videos with similar horizon positions. N6 used the \texttt{shape} lens to identify videos that contain people running. P4 used the \texttt{shape} lens to find videos containing circular shapes. P5 used the \texttt{shape} lens to find videos that have similar angles and are rotating in similar motions.

\textbf{Shine light on different perspectives.} Participants expressed that Swappable Lenses helped shine light on new perspectives (N1, N4, N5, P7): ``\textit{They give different perspectives that focus on different traits of the footage} (N1)''. For example, the \texttt{color} lens helped reveal videos with the same color palette or aesthetic (P1, P3, P4, P6): ``\textit{In the \texttt{color} lens, I could see the videos that have warmer colors versus cooler colors, which I didn't notice this in the \texttt{semantic} lens.} (P7)'' In addition, participants found the \texttt{shape} lens useful for finding videos containing people doing similar actions (N1, N5, N6, P5, P7) or videos with similar compositions (P4, P7). P7 felt that ``\textit{the different lenses created subtle filters to the footage which inspired [their] decision-making process.}''

\textbf{Something for everyone.} Participants expressed that Swappable Lenses allow VideoMap to be flexible in supporting editors who might prefer different types of editing techniques: ``\textit{It also serves users' preferences in different video editing techniques. Users can choose ways that work the best for them.} (N3)'' Among the 14 participants, 8 participants preferred using the \texttt{semantic} lens the most, 3 participants preferred using the \texttt{color} lens the most, and 3 participants preferred using the \texttt{shape} lens the most.

\subsubsection{Paths Explorer}

Participants were able to use Paths Explorer to identify suitable video transitions by recognizing interesting match cuts with minimal time and effort (RQ2). In addition, participants also expressed that Paths Explorer also helped encourage and facilitate more exploration. As part of the second task, we asked participants to record three video transitions they found the most interesting during their explorations. Some video transitions from participants include using the \texttt{shape} lens to cut between scenes that have buildings with similar architectural designs, using the \texttt{shape} lens to cut between scenes with similar composition (both having two prominent blocks of color divided at the horizon), using the \texttt{semantic} lens to cut between food scenes, and using the \texttt{color} lens to cut between scenes featuring trees with similar autumn tones.

\textbf{Identify match cuts.} Participants expressed that Paths Explorer is a ``\textit{cool and unique way to conceptually look at footage and generate match cut ideas} (P7)''. Participants were able to easily link together different environments with similar vibes (Principle 2) (P1, P7, N1, N2, N3, N7): ``\textit{[Paths Explorer] helps find associated concepts across different videos more easily.} (N2)'' Non-professional editors found that Paths Explorer changed how they think about video transitions. For example, N3 noted that Paths Explorer ``\textit{made [them] put more thought into transitions}'' where they ``\textit{previously had not thought too much about them}.'' N1 commented that Paths Explorer ``\textit{inspired a lot of new ideas about how a transition could be}.''

\textbf{Save time and effort.} Participants noted that Paths Explorer helped save time and effort in finding transitions (P3, P7, N1, N5): ``\textit{Scrolling over footage and going back and forth can be very time-consuming and tedious} (P7).'' P3 commented that ``\textit{having a lens that specifically found the paths between matching shapes shaved off hours of looking through footage}.'' N5 appreciated being able to see ``\textit{the exact frames that can be connected together}''.

\textbf{Encourage exploration.} Participants felt that Paths Explorer helped encourage more exploration compared to traditional video editing software (Principle 3): ``\textit{It's a more exploratory approach than with the Premiere interface. I can discover relationships between scenes easier. It's a great way to anticipate multiple possible transitions without committing to an idea.} (P4).'' N1 appreciated the ability to view multiple transition suggestions together: ``\textit{Oh these are kinda cool. This is probably better. I'll choose this one.} (N1)'' N2 enjoyed the ability to ``\textit{simultaneously explore frames at different time points and not being constrained by the temporal ordering of video content}.''

\subsubsection{Route Planner}

Participants were able to use Route Planner to quickly prototype rough cuts with automatic editing of clips (RQ3). In addition, participants also commented that Route Planner is able to help their editing workflows by generating ideas, facilitating quick previewing of ideas, and suggesting enhancements. As part of the third task, we asked participants to record a rough cut generated by Route Planner that they found the most interesting. Some rough cuts by participants include using the \texttt{shape} lens to generate a video of a person running with different backgrounds, using the \texttt{shape} lens to generate a video with a seamless foot transition, and using the \texttt{color} lens to generate a video that starts from daytime and ends with nighttime.

\textbf{Automatic editing.} Participants enjoyed being able to quickly generate edited rough cuts using Route Planner (P2, P3, N1, N2, N3, N6): ``\textit{It reduced the hassle of trying to arrange clips in a way that makes visual sense. Using traditional software would have made this process take hours.} (P3)'' P2 and N6 commented that Route Planner could be helpful for publishing video content on social media platforms: ``\textit{It could be great for platforms that need content generated quickly, like TikTok. It could easily help create content like a supercut or quick preview.} (P2)'' Participants felt that Route Planner generated quality cuts: ``\textit{It's a very cool video. That foot part didn't even seem like a transition. It just seemed so seamless that I thought it was the same video.} (N3)''

\textbf{Generate ideas.} Participants commented the Route Planner can be used for idea generation. For example, N1 commented that they could throw clips into Route Planner before jumping into editing to ``\textit{quickly plan out an editing plan.}'' On the other hand, N5 commented that they would primarily use Project Panel and Paths Explorer and ``\textit{use Route Planner if [they] run out of creative ideas}.''

\textbf{Preview ideas quickly.} Participants felt that Route Planner helps editors ``\textit{try out different ideas faster than the traditional software} (N6)'' (Principle 3). P7 commented that ``\textit{[Route Planner] was a really good way for [them] to visually come up with and execute ideas. When using Route Planner, I had thought of many different ways the videos and clips could have been edited together.}''

\textbf{Suggest enhancements.} Participants noted that Route Planner could be used as a tool for suggesting editing enhancements. P7 commented that they could first ``\textit{manually do a rough cut, then give Route Planner the clips that [they] used and see if it could come up with a better cut}.''

\begin{figure*}[!htb]
  \centering
  \includegraphics[width=12cm]{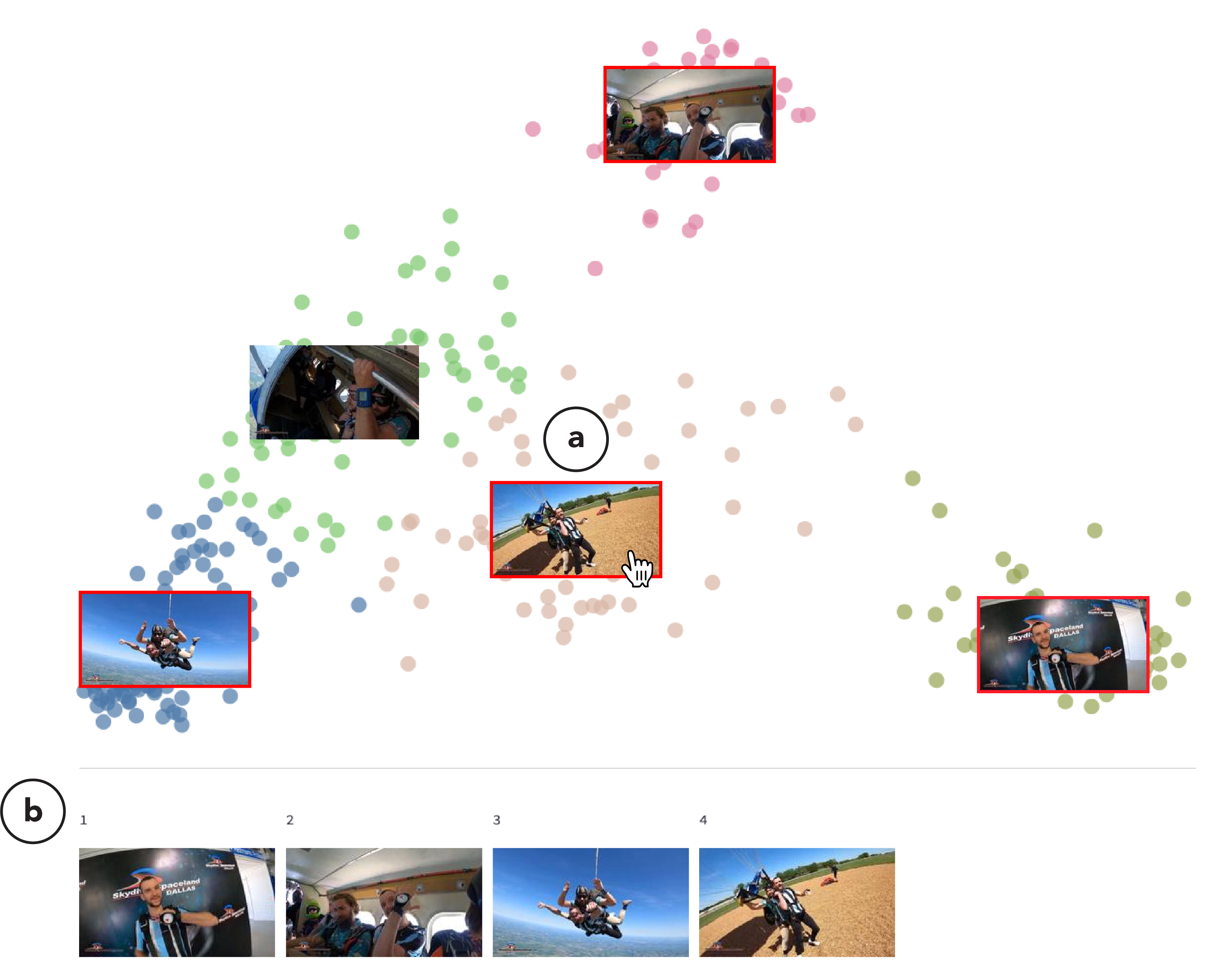}
  \caption{VideoMap's Project Panel can be extended to create summary videos. We automatically create ``semantic districts'' that approximately represent the main activities of a video using \textit{k}-means clustering under the \texttt{semantic} lens. The editor can select several landmarks to specify the activities to include in the summary video. Selections are highlighted with red borders (a) and displayed as a storyboard (b).}
  \Description{VideoMap's Project Panel can be extended to create summary videos. We automatically create ``semantic districts'' that approximately represent the main activities of a video using \textit{k}-means clustering under the \texttt{semantic} lens. The editor can select several landmarks to specify the activities to include in the summary video. Selections are highlighted with red borders (a) and displayed as a storyboard (b). }
  \label{fig:video-summarization}
\end{figure*}

\section{Discussion}
\label{section:discussion}

Participants generally felt that VideoMap offers a user-friendly approach to video editing (P3, P5, P6, N1, N2, N6). Specifically, non-professionals remarked that VideoMap ``\textit{makes editing tasks much easier} (N6)'', possesses a ``\textit{smoother learning curve} (N1)'', and offers a ``\textit{more fun way to edit videos} (N2).'' Additionally, participants expressed that VideoMap helps reduce the non-creative grunt work involved in video editing, allowing for greater focus on creative aspects (P5, N5, N6): ``\textit{Trying to write my process in Premiere made me realize how complicated it actually is to do this task, [and] how much manual searching and scrolling through frames I do.} (P5)''. N6 explained that their workflow of organizing video footage using traditional editors involves ``\textit{manually sorting the videos in a file browser, picking the videos that [they are] most interested in, then manually uploading them into the editing software} (N6)''. With Project Panel, they could ``\textit{have all videos uploaded at once, then filter with prompts using the ideas [they] have in mind} (N6)''.

One clear benefit of VideoMap is its enhanced overview capability (Principle 1). By breaking the sequential nature of browsing videos, viewing videos across a latent space in the Project Panel helps editors explore raw footage, enabling them to better familiarize themselves with and uncover structure within the footage (RQ1): ``\textit{[VideoMap] helped me get a more comprehensive overview of the materials more quickly} (P4)''. P7 reflected: ``\textit{I think the traditional video editor can leave much to be desired. [VideoMap] is a fresh take on the old concept, allowing for the clips to shine and visually show [me] what they are and how [I] might think to compose them.}'' N5 commented that ``\textit{[VideoMap] is a more inspirational way of organizing video clips}'' and N3 noted that ``\textit{[VideoMap] helps editors tell better stories with connected themes, color, and shape}''. Given an enhanced overview, participants were also able to visually ideate narratives, such as creating a video that weaves together one clip from each cluster on the map (P1, P5), or conversely, creating a video that only focuses on clips from one cluster (P6, N2). In addition, P5 was also able to utilize the overview to ``\textit{decide what other clips [they] need to take to complete the story}'', such as capturing more concert footage after noticing relatively small semantic concert clusters on the map.

By operating on the latent space, VideoMap helps visualize semantic or visual neighbors, thereby incentivizing editors to maintain continuity in their edits (Principle 2). Paths Explorer assists editors in brainstorming opportunities to create continuous transitions through transition recommendations to latent space neighbors (RQ2): ``\textit{[Paths Explorer] is a cool and unique way to conceptually look at footage and generate match cut ideas} (P7)''. N5 explained that ``\textit{using a traditional video editing software might involve jumping back and forth between the videos or overlaying frames on top of each other with a lower transparency to see if there are some common shapes}''. P3 contrasted editing transitions using Paths Explorer with their typical workflow: ``\textit{It really changes the process of editing. Usually, when you're doing this [with traditional editing software], you already have everything onto a timeline, and you're like, alright, now let's find the transitions.}'' Furthermore, non-professionals felt that Paths Explorer ``\textit{made [them] put more thought into transitions [where they] previously had not thought too much about them} (N3)'', and helped ``\textit{inspire a lot of new ideas about how a transition could be} (N1)''.

\begin{figure*}[!htb]
  \centering
  \includegraphics[width=11cm]{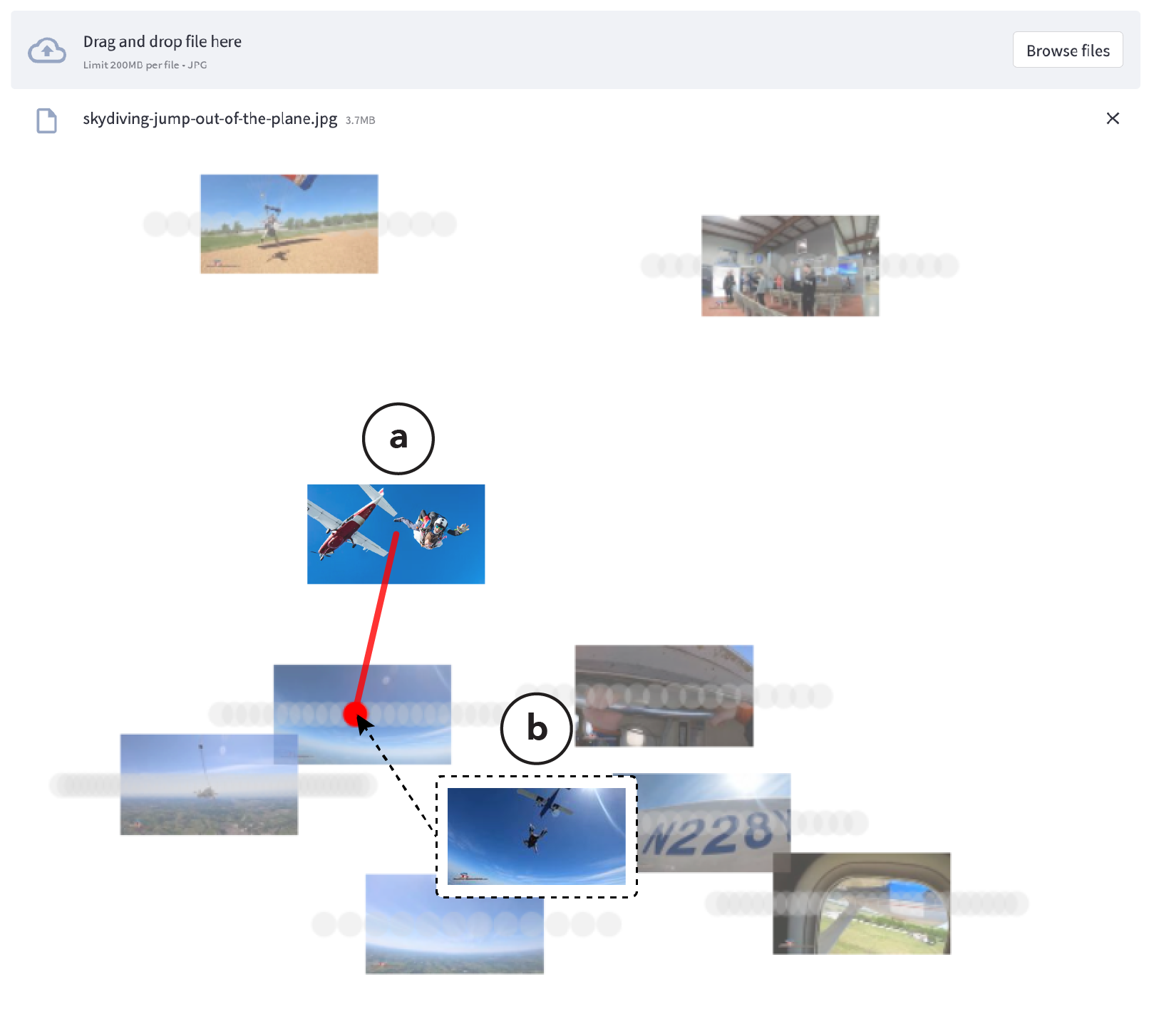}
  \caption{VideoMap's Paths Explorer can be extended to create highlight videos. The editor can upload a photograph (i.e., a custom landmark) depicting an activity (e.g., skydiving) (a). Our key insight is that photographs taken by photographers tend to capture the most highlight-worthy moments of an activity (e.g., when the skydiver jumps out of the aircraft). We then generate a highlight video using near neighbor video frames to the custom landmark in the \texttt{semantic} space (b).}
  \Description{VideoMap's Paths Explorer can be extended to create highlight videos. The editor can upload a photograph (i.e., a custom landmark) depicting an activity (e.g., skydiving) (a). Our key insight is that photographs taken by photographers tend to capture the most highlight-worthy moments of an activity (e.g., when the skydiver jumps out of the aircraft). We then generate a highlight video using near neighbor video frames to the custom landmark in the \texttt{semantic} space (b).}
  \label{fig:video-highlights}
\end{figure*}

VideoMap enables a more exploratory approach to editing through various components and mechanisms, such as automatic rough cuts with Route Planner, map-inspired navigation elements, prompt searching in Project Panel, and recommending multiple possible transitions in Paths Explorer (Principle 3). By automatically generating rough cuts, Route Planner helps editors quickly test and preview editing ideas (RQ3). Participants utilized the generated rough cuts for idea generation (``\textit{quickly plan out an editing plan} (N1)''), previewing the results of editing ideas (``\textit{I had thought of many different ways the videos and clips could have been edited together} (P7)''), and suggesting enhancements (``\textit{see if it could come up with a better cut} (P7)''). Participants commented positively on various interactions and map-inspired navigation elements, such as scrubbing landmarks to play through videos (P5, P6, N6) and selecting nodes to visualize paths to neighbors (P4, N1, N2). Participants enjoyed using prompts in the Project Panel to semantically search for clips in mind or discover new clips: ``\textit{It is like the difference between using Google Search and searching information from books} (N2).'' Additionally, P4 commented that ``\textit{Paths Explorer helped [them] discover and plan possible transitions much faster}'' and felt like a ``\textit{more exploratory approach than with [a traditional editing interface]}'', given the ability ``\textit{to anticipate multiple possible transitions without committing to an idea}''. 

Finally, we envision a vast and versatile design space for VideoMap, with future developers enhancing and extending its capabilities. Our current proof-of-concept enables editors to edit video through notions of \texttt{semantics}, \texttt{color}, and \texttt{shape}. We plan to open-source our code so that developers can build new lenses to support additional editing tasks. Furthermore, we hope that VideoMap can be used as a building block to support the development of new creative video editing applications. In the following section, we demonstrate some examples.

\begin{figure*}[!htb]
  \centering
  \includegraphics[width=15cm]{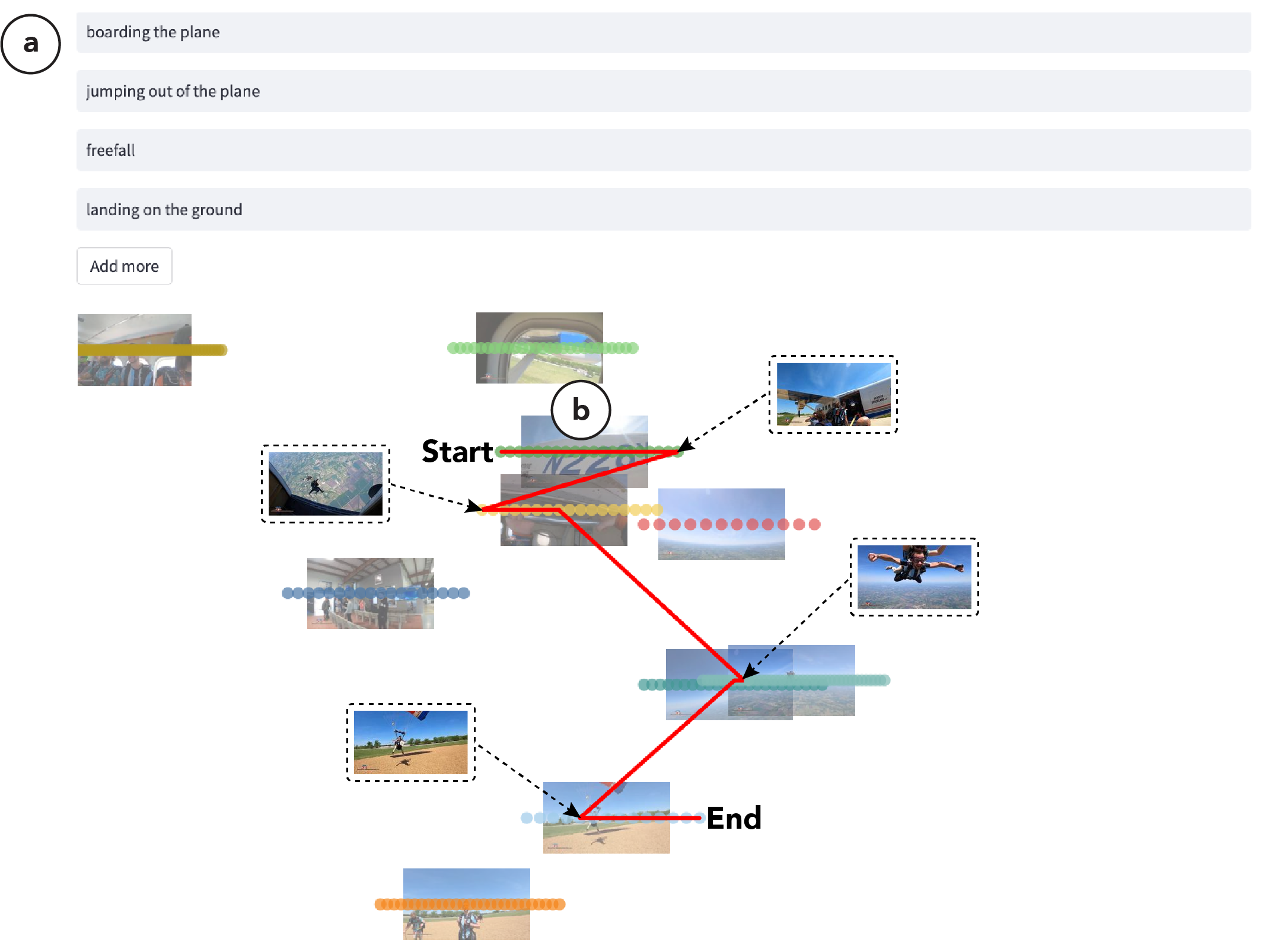}
  \caption{VideoMap's Route Planner can be extended to edit videos using text. The editor can describe a desired video using descriptive sentences, like writing a story (a). We then match each sentence to the closest video clip in the \texttt{semantic} space and generate a video by finding the shortest route along the clips (b).}
  \Description{VideoMap's Route Planner can be extended to edit videos using text. The editor can describe a desired video using descriptive sentences, like writing a story (a). We then match each sentence to the closest video clip in the \texttt{semantic} space and generate a video by finding the shortest route along the clips (b).}
  \label{fig:text-based-video-editing}
\end{figure*}

\section{Extended Applications}

We implemented three extended applications of VideoMap to demonstrate how it can be customized for additional use cases. These applications include (1) video summarization with semantic districts, (2) video highlights with custom landmarks, and (3) text-based video editing.

\subsection{Video Summarization}
\label{section:video-summarization}

We show how VideoMap Project Panel can be extended to help editors create summary videos (Figure \ref{fig:video-summarization}). We first project the source video's frames into a \texttt{semantic} latent space. Instead of creating districts per video, we apply \textit{k}-means clustering to automatically partition the video frames into ``semantic districts.'' Since we are working in a semantic space, each district approximately represents a main activity of the source video. We determine the number of districts (i.e., \textit{k} value) using the elbow method \cite{thorndike1953belongs}. We color-code the districts and visualize the centroid frame of each district as semantic landmarks. To create the summary video, the editor can add semantic districts (i.e., main activities) by clicking on the landmarks, in the order they want them to appear. Selected landmarks are highlighted with red borders (Figure \ref{fig:video-summarization}a) and displayed as a storyboard at the bottom (Figure \ref{fig:video-summarization}b). We generate the summary video by taking three seconds of video centered around the landmark for each semantic district).

\subsection{Video Highlights}
\label{section:video-highlights}

We show how VideoMap Paths Explorer can be extended to help editors create highlight videos (Figure \ref{fig:video-highlights}). The editor can first upload a photograph depicting an activity (e.g., skydiving). We then project the photograph into the latent space under the \texttt{semantic} lens and visualize it as a ``custom landmark'' (Figure \ref{fig:video-highlights}a). Next, we draw a path from the custom landmark to the nearest video frame in the latent space (Figure \ref{fig:video-highlights}b). Videogenic \cite{lin2024videogenic} found that photographs taken by photographers tend to capture the most highlight-worthy moments of an activity (e.g., when the skydiver jumps out of the aircraft). Thus, we take the photograph's neighboring video frames as the highlight video. We generate the highlight video by taking five seconds of video centered around the nearest video frame).

\subsection{Text-Based Video Editing}
\label{section:text-based-video-editing}

We show how VideoMap Route Planner can be extended to support editors in text-based video editing (Figure \ref{fig:text-based-video-editing}). The editor can first describe the video they want to create using descriptive sentences, like writing a story (Figure \ref{fig:text-based-video-editing}a). For each sentence, we match the closest video in the semantic space, using the same CLIP-based method for comparing text and image embeddings as detailed in our implementation for Prompts (see Section \ref{section:prompts}). We then create the shortest route along the videos using the same dynamic programming approach as detailed in Route Planner (see Section \ref{section:route-planner}) with an additional ordering constraint (i.e., in the order of the story) (Figure \ref{fig:text-based-video-editing}b).

\section{Future Work}
\label{section:future-work}

While VideoMap was positively received in our user study, there are several avenues for improvement that we plan to address for future work. 
First, VideoMap currently supports three types of lenses that facilitate the formation of narratives (\texttt{semantic}) and the creation of visual transitions (\texttt{shape} and \texttt{color}). In our study, participants suggested additional lenses, such as a cinematography lens to cluster videos by different shot types (e.g., by training a shot type classification model), an Image Signal Processor (ISP) lens to cluster videos by white balance, focus, and exposure values (e.g., by utilizing image analysis algorithms), and a motion lens to cluster videos by distinct camera movements (e.g., by computing motion vectors with optical flow \cite{lucas1981iterative}). For future work, we plan to open-source the pipeline code for developers to build new lenses by generating new latent spaces that are meaningful for video editing. This pipeline could involve generating embeddings using \textit{AI foundation models} \cite{bommasani2021opportunities} or training custom models tailored to specific video editing objectives. 
Moreover, we could also explore the creation of ``hybrid lenses.'' For instance, we could create a lens with weightings of 80\% shape and 20\% semantics to enable the discovery of shape-matching cuts that also possess related semantics. Hybrid lenses could provide editors with the flexibility of blending together the properties of multiple lenses, allowing them to fine-tune the latent space according to their editing goals.
\camera{Second, due to time constraints, the current design of our user study consists of users trying out VideoMap without a comparative baseline system. For future work, it may be interesting to compare editing using VideoMap against editing on traditional sequential editing timelines and analyze how VideoMap could change the way people engage in video editing.}
\camera{Finally, while VideoMap currently caters to primarily maintaining the continuity principle in video editing, sometimes editors may wish to intentionally break this principle and connect two contrasting scenes to create a shocking effect. To support this, we may extend the Paths Explorer to find far-away cuts. Being able to find contrasting yet sensible transitions could also be an interesting exploration area.}

% Finally, VideoMap currently creates latent spaces of video frames. We could explore extending VideoMap to incorporate audio by projecting both video frames and audio into a joint multimodal latent space, enabling editors to also edit sound using VideoMap.

\section{Conclusion}

\camera{This research explores a new paradigm of video editing based on latent space representations of video frames. We distilled a set of principles grounded on editing practices and theories and built VideoMap, a proof-of-concept latent space video editing interface.} We facilitate navigation of the latent space through map-inspired navigational elements and enable switching between different latent spaces using swappable lenses. We implemented three VideoMap components to assist editors in three editing tasks: exploring video footage, brainstorming suitable video transitions, and rapidly prototyping rough cuts. In a user study with both professionals and non-professionals, we found that VideoMap provides a user-friendly editing experience, reduces tedious grunt work, enhances the overview capability of video footage, promotes editing continuity, and enables a more exploratory approach to video editing. We implemented three extended applications to demonstrate how future developers may customize VideoMap for additional editing use cases.
\camera{Recently, there has been an explosion of diverse AI foundation models \cite{foundation-models} capable of encoding video data into a variety of latent space representations (semantics, 3D depth, human poses, audio, …) \cite{lin2024jigsaw}. We hope VideoMap’s capabilities can be extended over time through the creation of new AI-powered lenses and support creative ways of co-creating video content with AI on a latent space interface.}

%%
%% The acknowledgments section is defined using the "acks" environment
%% (and NOT an unnumbered section). This ensures the proper
%% identification of the section in the article metadata, and the
%% consistent spelling of the heading.

\begin{acks}
\rev{We would like to thank colleagues in the Augmented Design Capability Studio for providing valuable feedback on our system. We would like to thank Howie Wang for the idea of remapping video frames of the same video chronologically. We would like to thank Bella Yang and Bill Guo for filming video clips used for the user study. We would like to thank Dominik Moritz, Thiago Teixeira, and Lukas Masuch for support on the Streamlit Vega-Lite package used in VideoMap's implementation.}
\end{acks}

%%
%% The next two lines define the bibliography style to be used, and
%% the bibliography file.
\bibliographystyle{ACM-Reference-Format}
\bibliography{main}

%%
%% If your work has an appendix, this is the place to put it.

\raggedbottom
\pagebreak

\appendix

\end{document}